\documentclass[linenumbers, twocolumn]{aastex631}

\usepackage{amsmath}

\newcommand{\totalsources}{200 }

\newcommand{\caldensity}{1 per 100 square degrees}
\newcommand{\caldensityfov}{2 in-beam calibrators per field of view}

\newcommand{\caldensitydeep}{1 per 20 square degrees}
\newcommand{\caldensitydeepfov}{14 in-beam calibrators per field of view}

\newcommand{\threshold}{100 mJy}

\usepackage{siunitx}
\DeclareSIUnit{\parsec}{pc}
\DeclareSIUnit{\dmunit}{pc cm^{-3}}
\DeclareSIUnit{\jansky}{Jy}

\received{\today}
\revised{\today}

\submitjournal{AJ}

\shorttitle{VLBI Calibrator Grid at 600\,MHz}
\shortauthors{Andrew et al.}
\graphicspath{{./}{figures/}}

\begin{document}
\nolinenumbers
\title{A VLBI Calibrator Grid at 600\,MHz for Fast Radio Transient Localizations with CHIME/FRB Outriggers}
\author[0000-0002-3980-815X]{Shion Andrew}
\affiliation{Massachusetts Institute of Technology, 77 Massachusetts Ave, Cambridge, MA 02139, USA}
\affiliation{Department of Physics, Massachusetts Institute of Technology, 77 Massachusetts Ave, Cambridge, MA 02139, USA}
 
\author[0000-0002-4209-7408]{Calvin Leung}
  \affiliation{Department of Astronomy, University of California Berkeley, Berkeley, CA 94720, USA}
  \altaffiliation{NASA Hubble Fellowship Program~(NHFP) \\ and Einstein Fellow}

\author[0009-0000-2762-5957]{Alexander Li}
\affiliation{Department of Physics, Massachusetts Institute of Technology, 77 Massachusetts Ave, Cambridge, MA 02139, USA}
 
\author[0000-0002-4279-6946]{Kiyoshi W. Masui}
\affiliation{Massachusetts Institute of Technology, 77 Massachusetts Ave, Cambridge, MA 02139, USA}
\affiliation{Department of Physics, Massachusetts Institute of Technology, 77 Massachusetts Ave, Cambridge, MA 02139, USA}

\author[0000-0001-5908-3152]{Bridget C. Andersen}
\affiliation{Department of Physics, McGill University, 3600 rue University, Montr\'eal, QC H3A 2T8, Canada}
\affiliation{Trottier Space Institute, McGill University, 3550 rue University, Montr\'eal, QC H3A 2A7, Canada}

\author[0000-0003-3772-2798]{Kevin Bandura}
  \affiliation{Lane Department of Computer Science and Electrical Engineering, 1220 Evansdale Drive, PO Box 6109, Morgantown, WV 26506, USA}
  \affiliation{Center for Gravitational Waves and Cosmology, West Virginia University, Chestnut Ridge Research Building, Morgantown, WV 26505, USA}
  
\author[0000-0002-8376-1563]{Alice P. Curtin}
  \affiliation{Department of Physics, McGill University, 3600 rue University, Montr\'eal, QC H3A 2T8, Canada}
\affiliation{Trottier Space Institute, McGill University, 3550 rue University, Montr\'eal, QC H3A 2A7, Canada}

\author[0000-0003-4810-7803]{Jane Kaczmarek}
\affiliation{CSIRO Space \& Astronomy, Parkes Observatory, P.O. Box 276, Parkes NSW 2870, Australia}

\author[0000-0003-2116-3573]{Adam E. Lanman}
\affiliation{Massachusetts Institute of Technology, 77 Massachusetts Ave, Cambridge, MA 02139, USA}
\affiliation{Department of Physics, Massachusetts Institute of Technology, 77 Massachusetts Ave, Cambridge, MA 02139, USA}

\author[0000-0002-5857-4264]{Mattias Lazda}
  \affiliation{Dunlap Institute for Astronomy \& Astrophysics, University of Toronto, 50 St.~George Street, Toronto, ON M5S 3H4, Canada}
  \affiliation{David A.~Dunlap Department of Astronomy \& Astrophysics, University of Toronto, 50 St.~George Street, Toronto, ON M5S 3H4, Canada}

\author[0000-0002-0772-9326]{Juan Mena-Parra}
  \affiliation{Dunlap Institute for Astronomy \& Astrophysics, University of Toronto, 50 St.~George Street, Toronto, ON M5S 3H4, Canada}
  \affiliation{David A.~Dunlap Department of Astronomy \& Astrophysics, University of Toronto, 50 St.~George Street, Toronto, ON M5S 3H4, Canada}

\author[0000-0002-2551-7554]{Daniele  Michilli }
\affiliation{MIT Kavli Institute for Astrophysics and Space Research, Massachusetts Institute of Technology, 77 Massachusetts Ave, Cambridge, MA 02139, USA}
\affiliation{Department of Physics, Massachusetts Institute of Technology, 77 Massachusetts Ave, Cambridge, MA 02139, USA}

\author[0000-0003-0510-0740]{Kenzie Nimmo}
  \affiliation{MIT Kavli Institute for Astrophysics and Space Research, Massachusetts Institute of Technology, 77 Massachusetts Ave, Cambridge, MA 02139, USA}
  
\author[0000-0002-8912-0732]{Aaron~B.~Pearlman}
  \altaffiliation{Banting Fellow, McGill Space Institute~(MSI) Fellow, \\ and FRQNT~Postdoctoral Fellow.}
  \affiliation{Department of Physics, McGill University, 3600 rue University, Montr\'eal, QC H3A 2T8, Canada}
  \affiliation{Trottier Space Institute, McGill University, 3550 rue University, Montr\'eal, QC H3A 2A7, Canada}
  
\author[0000-0003-1842-6096]{Mubdi Rahman}
  \affiliation{Sidrat Research, 124 Merton Street, Suite 507, Toronto, ON M4S 2Z2, Canada}

 \author[0000-0002-4823-1946]{Vishwangi Shah}
  \affiliation{Department of Physics, McGill University, 3600 rue University, Montr\'eal, QC H3A 2T8, Canada}
  \affiliation{Trottier Space Institute, McGill University, 3550 rue University, Montr\'eal, QC H3A 2A7, Canada}

\author[0000-0002-6823-2073]{Kaitlyn Shin}
  \affiliation{MIT Kavli Institute for Astrophysics and Space Research, Massachusetts Institute of Technology, 77 Massachusetts Ave, Cambridge, MA 02139, USA}
  \affiliation{Department of Physics, Massachusetts Institute of Technology, 77 Massachusetts Ave, Cambridge, MA 02139, USA}
 
\author[0000-0002-1491-3738]{Haochen Wang}
\affiliation{Massachusetts Institute of Technology, 77 Massachusetts Ave, Cambridge, MA 02139, USA}
\affiliation{Department of Physics, Massachusetts Institute of Technology, 77 Massachusetts Ave, Cambridge, MA 02139, USA}

\correspondingauthor{Shion Andrew}
\email{shiona@mit.edu}




\collaboration{99}{(CHIME/FRB Collaboration)}



\begin{abstract}
\nolinenumbers
The Canadian Hydrogen Intensity Mapping Experiment Fast Radio Burst (CHIME/FRB) Project has a new VLBI Outrigger at the Green Bank Observatory (GBO), which forms a 3300\,km baseline with CHIME operating at 400--800\,MHz. Using 100\,ms long full-array baseband ``snapshots'' collected commensally during FRB and pulsar triggers, we perform a shallow, wide-area VLBI survey covering a significant fraction of the Northern sky targeted at the positions of compact sources from the Radio Fundamental Catalog. In addition, our survey contains calibrators detected from two 1s long trial baseband snapshots for a deeper survey with CHIME and GBO. In this paper, we present the largest catalog of compact calibrators suitable for 30-milliarcsecond-scale VLBI observations at sub-GHz frequencies to date. Our catalog consists of \totalsources total calibrators in the Northern Hemisphere that are compact on 30-milliarcsecond scales with fluxes above \threshold. This calibrator grid will enable the precise localization of hundreds of FRBs a year with CHIME/FRB-Outriggers. 

\end{abstract}



\section{Introduction}
\label{sec:intro}

A peculiarity of radio surveys is that the sky looks very different at different angular resolutions. Sources which appear bright in one survey can be undetectable at high angular resolutions, and vice versa. This is because radio surveys probe a highly specific range of angular scales: while single dish telescopes are sensitive to spatially-extended emission such as H II regions and Galactic synchrotron radiation, the most extended interferometers are sensitive to compact, point-like emission such as pulsars, ordinary stars, and active galactic nuclei (AGN). 

A complete understanding of the radio sky requires surveys probing a wide range of angular scales. Very long baseline interferometry (VLBI) surveys represent one extreme end of this spectrum, offering valuable information at the finest angular scales possible in modern astronomy. VLBI surveys are also rarer in number largely due to a much smaller average field of view (historically $\sim$ arcseconds) which limits the rate at which the sky can be surveyed. While generations of point source surveys (\cite{Preston_1985}; see~\citet{Petrov_2021} and references therein for a modern perspective) have expanded the sample of compact sources used by the VLBA as calibrators at 1.5\,GHz and above, these calibrators are not directly transferable to lower frequency ($<$1\,GHz) VLBI observations due to low frequency turnover (e.g. synchrotron self-absorption) and increased angular broadening. Because a wide-area VLBI calibrator survey at sub-GHz frequencies has not yet been undertaken, VLBI at these frequencies remains relatively unexplored.


There are numerous observational opportunities that could be uniquely enabled by low-frequency VLBI. This includes angular broadening from interstellar scattering \citep{angular_broadening_1989}, astrometric measurements of steep spectrum pulsars \citep{steep_spectrum_pulsar_2021}, and more recently, the precise
localization of fast radio bursts (FRBs) with upcoming VLBI survey instruments operating at sub-GHz frequencies such as CHIME/FRB Outriggers \citep{kko_adam} and CHORD \citep{Vanderlinde_2019}. Yet in the past several decades, only a few attempts have been made to systematically survey the sky for VLBI calibrators at $<1$\,GHz. Notably, the Long Baseline Calibrator Survey~\citep{moldon2015lofar,jackson2016lbcs,Jackson_2022,Morabito_2022}, a dedicated survey using the international baselines ($\sim 1100$ km) of LOFAR, reported $>$20,000 calibrators for the LOFAR array. However, the LBCS only probes 0.5\,M$\lambda$ angular scales, leaving a gap in the wavelength versus compactness phase space (see Fig.~\ref{fig:megalambda_freq_space}). 

Other sub-GHz frequency VLBI surveys at $>$1\,M$\lambda$ angular scales have been limited in number. ~\citet{Clark_1975} observed $\approx 20$ sources brighter than 1 Jy at 96-197\,MHz on a 2600 km baseline, and ~\citet{Chuprikov_1999} observed 5 unresolved sources at 327\,MHz on a 2800 km baseline. ~\citet{lenc2008deep} discovered a total of 19 sources with the European VLBI Network, but over half of these were on the edge of the fields of view where the effective angular resolution is reduced to $0.5$\,M$\lambda$. Finally, in anticipation of the construction of Low Frequency Array (LOFAR),~\citet{Rampadarath_2009} presented a comprehensive re-analysis of 44 archival VLBA 327\,MHz observations prior to 2009 at varying angular resolutions (2-9\,M$\lambda$), detecting $15$ sources brighter than 1 Jy. The net result of these efforts is that over the whole sky, less than a hundred sources have directly been confirmed to possess bright compact structure at $<1$\,GHz on the $2-8$\,M$\lambda$ scales relevant for VLBI imaging.

The primary goal of this work is to help fill in this gap in frequency-compactness phase space for existing VLBI calibrators and provide a calibrator grid needed for the CHIME/FRB Outriggers project, which aims to localize a substantial sample of FRBs to 50-milliarcsecond precision upon detection. Using CHIME and the Green Bank Outrigger (GBO)–the longest of the CHIME/FRB-Outrigger baselines at 3300\,km–we performed commensal observations of targets in the VLBA Calibrator Grid. Our detections provide a quasi-uniform grid of low-frequency fringe finders in the northern ($\delta > 0^\circ$) sky, consisting of the brightest ($>\SI{100}{\milli\jansky}$) and most compact ($\sim 0.03'' \leftrightarrow 6\,$M$\lambda$) sources in the sky.


The rest of this paper proceeds as follows. The survey design and search strategy is described in Section \ref{sec:observation_strategy}. Our catalog, its statistics, and cross-matches in Section \ref{sec:catalog}. Applications for the sources in our catalog are discussed in Section \ref{sec:Applications}.

\section{Observational Strategy with CHIME \& GBO} 
\label{sec:observation_strategy}

\subsection{Integration times and sensitivity}
CHIME and its Outriggers are each wide-field interferometers operating from 400-800\,Mhz with a primary beam FWHM of $\sim$ 90$^\circ$ (N--S) x 2$^\circ$ (E--W) \citep[CHIME Collaboration et al. 2024 in prep]{chime_overview,kko_adam,chime_frb_2018}. Each of the Outrigger cylinders are rotated and rolled to share the same field of view as CHIME. Upon detection of a radio transient, CHIME sends triggers to all its Outriggers to record the baseband (raw voltage) data encompassing the burst \citep{chime_frb_2018}. The baseband data contains 1024 channels each with a spectral resolution of 390\,kHz (2.56\,$\mu$s sample rate). The widefield nature of these full-array baseband ``snapshots'' make it an extremely efficient way to survey wide areas. Due to real-time data rate limitations, data can be recorded for a maximum of $1.4$\,s, and during ordinary operations, the snapshot duration is only $\sim 100$\,ms. 

The majority of our calibrator search was conducted commensally on all $\sim 100$\,ms  baseband snapshots recorded from November 2023 through July 2024. Since the system equivalent flux density (SEFD) of CHIME per formed beam is $\sim 30$\,Jy, and at GBO is $\sim 240$\,Jy, a 100\,mJy steady radio source can be detected with a signal to noise (S/N) of 10 in a $100$\,ms integration. Additionally, we collected two 1.4\,s long baseband snapshots in October 2024 and December 2024, to probe the calibrator density for sources brighter than $\sim$ 10mJy. These snapshots provide a pilot study for a future, deeper calibrator survey to obtain a denser grid of available calibrators for CHIME/FRB-Outriggers in the sky. Prospects for a future survey with CHIME/FRB Outriggers is discussed in more detail in Section \ref{sec:deeper}.

%
\begin{figure}
    \centering
    \includegraphics[width=\linewidth]{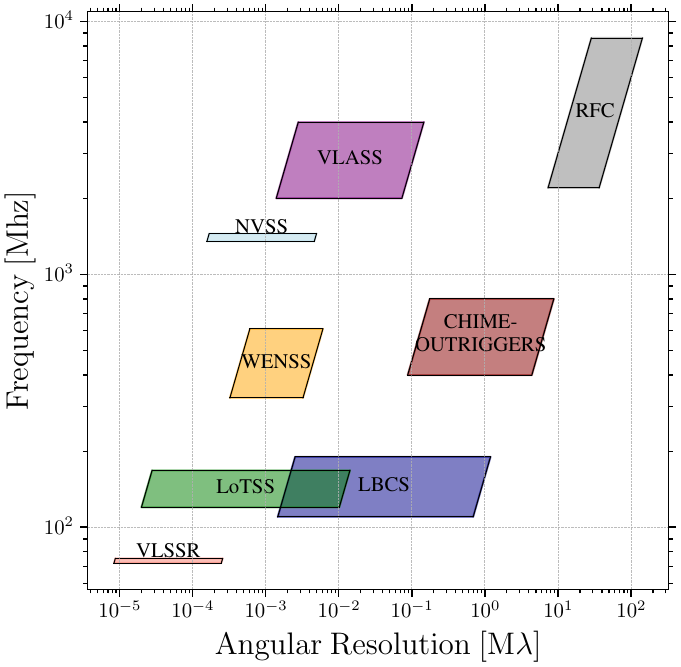}
    \caption{Frequency versus angular resolution for the VLA Sky Survey (VLASS) \citep{Lacy_2020}, NRAO VLA Sky Survey (NVSS) \citep{Condon_1998}, the Westerbork Northern Sky Survey (WENSS) \citep{Rengelink_1997}, the VLA Low-Frequency Sky Survey (VLSSR) \citep{lane_2014}, LOFAR Two-metre Sky Survey (LotSS) \citep{Shimwell_2017}, the Lofar Long Baseline Calibrator Survey (LBCS) \cite{jackson2016lbcs}, and the Radio Fundamental Catalog (RFC). The frequency and angular resolution range for the CHIME/FRB-Outriggers array is included in red.}
    \label{fig:megalambda_freq_space}
\end{figure}


\subsection{Target Selection \& Field of View}
 The common field of view of CHIME and its VLBI Outriggers covers over 200 square degrees, allowing access to potentially hundreds of in-beam VLBI calibrators at any given moment in time
\citep[see Figure 5]{leung2020synoptic}. However, imaging this huge area would be extremely inefficient due to the sparseness of the radio sky, the limited u-v coverage of our array, and the angular resolution of our array, which is roughly 10 milliarcseconds. Therefore for each snapshot, we targeted all sources from the Radio Fundamental Catalog (RFC) \footnote{ \url{https://astrogeo.org/sol/rfc/rfc_2024b/}} that were within the instantaneous field of view of our array, which we define to be the 25$\%$ sensitivity level of the CHIME primary beam: $\pm$ 3$^\circ$ East-West and $\pm$ 60$^\circ$ North-South of the local zenith angle at CHIME. 

While other radio catalogues could have been included in our survey (e.g. see Fig.~\ref{fig:megalambda_freq_space}), our primary consideration for target selection was the availability of absolute astrometric positions. CHIME/FRB Outriggers aims to localize FRBs to 50 milliarcsecond precision; thus for astrometric VLBI with CHIME/FRB Outriggers, the calibrator positions must be known to at least 50 milliarcseconds. We note that the availability of absolute astrometry is easily met by the RFC but not by the LBCS, which does not measure astrometric positions directly and instead reports astrometric positions from the WENSS survey. 

We also note that one important frequency-dependent effect on VLBI astrometry is the so-called ``core shift,'' an effect widely present in jet emission~\citep{Plavin_2019} in which the observed position of an AGN core shifts as a function of frequency due to synchrotron self-absorption (see e.g.~\citet{Pushkarev_2012}). However, these effects are measured to be relevent on $\sim 1$ mas scales \citep{Solokovsky_2011}, which is below the astrometric systematics floor for CHIME/FRB-Outriggers (50 mas).

\subsection{Post Correlator Search Strategy}
\label{sec:search_strategy}
We used multiple phase center beamforming to efficiently form station beams towards each calibrator position. After beamforming the 1024 and 128 dual-linear-polarization inputs at CHIME and GBO, respectively, the phased-array data are written to disk. Then we use the PyFX correlator~\citep{Leung_2024} to correlate the phased-array data, producing visibilities for each pointing over the full 400--800\,MHz bandwidth. 

Up to $\sim 40\%$ of CHIME's band can be corrupted completely by RFI, resulting in a $\sim 20\%$ drop in sensitivity for days with particularly bad RFI environment. RFI contaminated channels were identified with a spectral kurtosis estimator ($\hat{SK}$) where all channels with a $\hat{SK}$ value 3 standard deviations away from the mean were removed (see \cite{Nita_2010}, \cite{Taylor_2019} for the full definition of $\hat{SK}$). 

After RFI is removed, we define our VLBI signal-to-noise ratio as the amplitude of our signal fringe at delay $\tau(\hat{n})$ (for a source at sky position $\hat{n}$) relative to the fluctuation of noise fringes after performing a Fourier transform over frequency:
\begin{equation}
    \rm{S/N}=\frac{|\rm{MAX}[\tilde{V}(\tau(\hat{n}))]|}{\sigma_\tau}
\end{equation}
where
\begin{equation}
    \tilde{V}(\tau(\hat{n}))=\mathcal{F}[V(\nu)](\tau(\hat{n}))
\end{equation}
denotes the Fourier transformed visibilities in the delay domain, which are maximized at the delay corresponding to the target $\tau(\hat{n}_{\rm{target}})$ for a detection, and 
\begin{equation}
    \sigma_\tau^2= \rm{Var}[\tilde{V}]_{\tau}
\end{equation}
is the variance of the visibilities over a range of delays in the range of $\pm 1.28 \mu s$. 

One subtlety of this S/N metric is that the ionosphere, which introduces quadratic and higher order terms in the phase delay as a function of frequency, will degrade the coherent S/N if those terms are not removed.  We therefore maximize the coherent S/N metric over a range of trial ionospheric slant total electron content (sTEC) values ranging from $\pm$50 TECu. This fits out the ionosphere, which is equivalent to conducting a two-dimensional S/N-maximizing search over both delay and sTEC as done in other low-frequency VLBI observations, e.g.~\citet{vanweeren2016facet,sanghavi2023tone}.


After maximizing the S/N for a particular pointing, we define a detection as a coherent S/N exceeding 10 on the CHIME-GBO baseline in either parallel hand polarizations (XX or YY). While the compact radio sky is relatively sparse, we expect occasional ``spurious" signals due to (for example) side lobe detections or source confusion. 

Our detections were flagged as ``spurious" if the fringe was found in the cross-correlated visibilities at a relative delay larger than half an integer frame ($\pm 1.28 \mu s$); this delay window is approximately equivalent to an angular distance of 13" on the CHIME-GBO baseline. For sources that did not contain another in-beam calibrator (for phase referencing), this is the maximum delay expected from relative clock and cable delays between sites. We found a total of three such spurious sources in our survey, which we exclude in our final catalog.

\section{Source Catalog}
\label{sec:catalog}
\subsection{Calibrator Density}
\begin{figure*}[t]
    \centering
    \includegraphics[width=\textwidth]{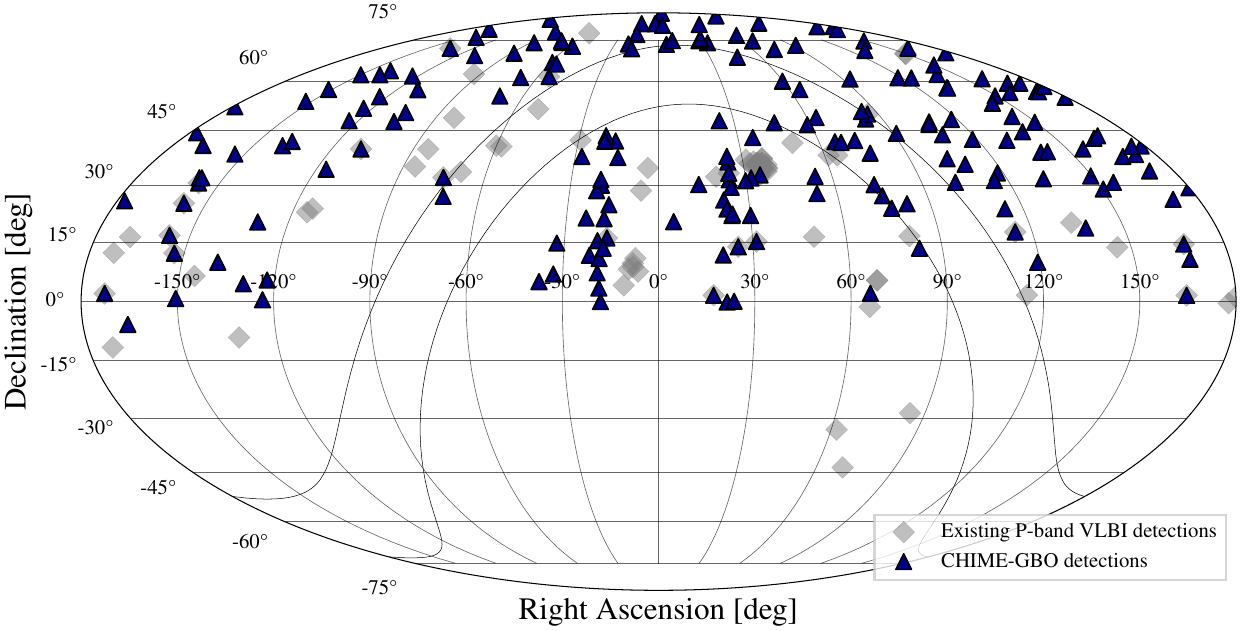}
    \caption{All the sources detected in our survey, with the Galactic plane shown at $b=\pm10$º. The 44 existing P-band VLBI detections of sources above a threshold of 1 Jy were compiled from \citep{Rampadarath_2009,Clark_1975,Chuprikov_1999}, as well as~\citep{lenc2008deep}. This shows that our preliminary fringe-finder catalog significantly extends the current list of P-band fringe finders to a substantial fraction of the sky.}
    \label{fig:calibrator_mollview}
\end{figure*}

Figure \ref{fig:calibrator_mollview} shows the distribution of all \totalsources sources detected in our survey. Excluding the Galactic plane–where we expect a strong negative selection bias due to angular broadening, the average calibrator density we observe at the $\sim 1$ Jy completeness threshold is \caldensity \ on 0.03'' scales, or approximately \caldensityfov. 

\begin{figure}
    \centering
    \includegraphics[width=\linewidth]{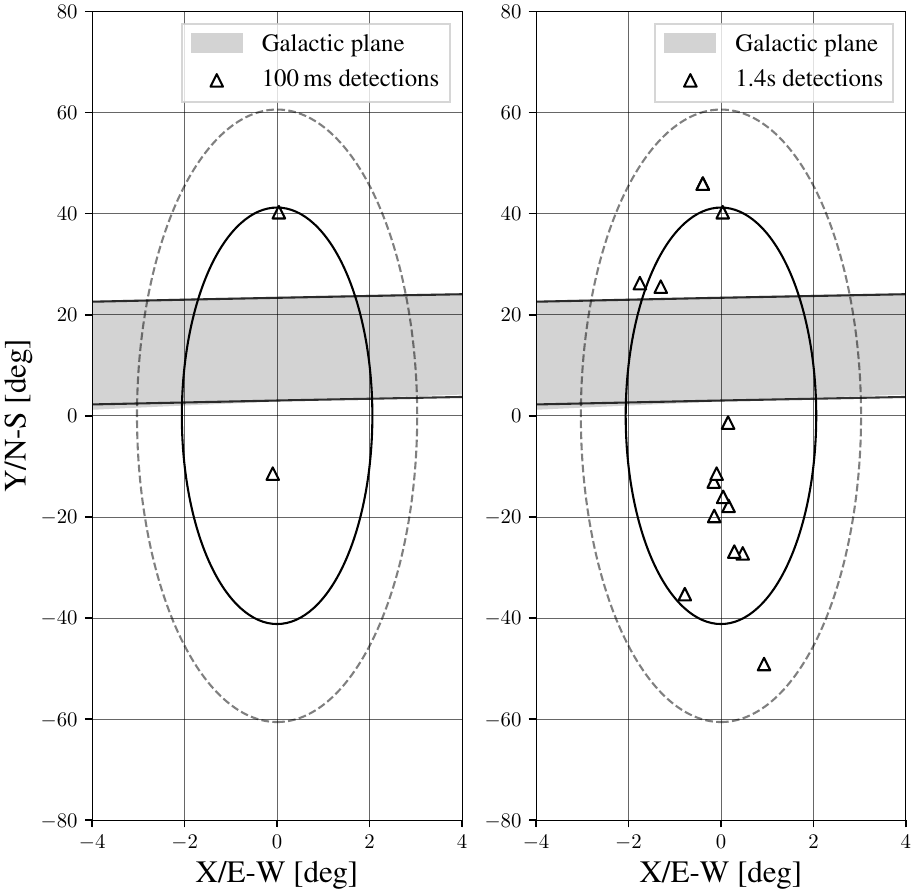}
    \caption{Increase in calibrator density for a given FOV from 100ms integration (left) to 1.4\,s integration (right). The band-averaged CHIME beam at 50\% and 25$\%$ sensitivity is shown in the gray and black contours, respectively. The Galactic plane ($|b|<10^\circ$) is overlaid in gray. This illustrates that conducting a deeper survey with 1.4\,s snapshots will likely densify the calibrator grid by a factor of $\sim$7.}
    \label{fig:deeper_integration_fov}
\end{figure}

As shown in Figure \ref{fig:deeper_integration_fov}, the average calibrator density over our two 1.4\,s integration snapshots increases to \caldensitydeep, or approximately \caldensitydeepfov. Our flux threshold for detection scales with integration time $t$ as 
\begin{equation}
    F_{\rm{thresh}}\propto t^{-1/2}
\end{equation}
or equivalently the distance threshold for a source of fixed intrinsic luminosity scales as
\begin{equation}
    d_{\rm{thresh}}\propto t^{1/4}.
\end{equation}
If we assume these radio sources to be uniformly distributed in an Euclidean universe ($N_{\rm{sources}}\propto d^3$), then we expect
\begin{equation}
    N_{\rm{detections}}\propto t^{3/4}
\end{equation}
which is consistent with the increase in calibrator density we observe in our 1.4\,s integration snapshots. Thus these trial snapshots indicate that a future survey that goes even moderately deeper can significantly help densify our current calibrator grid. 



\section{Applications}
\label{sec:Applications}
\subsection{Future Prospects for a Deeper Survey }
\label{sec:deeper}
The most direct application of our survey is as astrometric calibrators for future low frequency VLBI telescopes. In particular, it will enable the VLBI capabilities for CHIME/Outriggers at 400--800\,MHz, which in conjunction with the CHIME FRB backend will be able to localize hundreds of FRBs a year to $\sim$tens of milliarcseconds to better understand their astrophysical origins. In addition to the upcoming CHIME/FRB Outriggers project, several planned FRB facilities~\citep{Vanderlinde_2019,Lin_2022} will also rely on VLBI Outriggers at $\sim 600$\,MHz for the localization of fast radio bursts and other transients.

The results from our current 1.4\,s snapshots in Section \ref{sec:catalog} indicate that a future calibrator survey with CHIME-GBO consisting entirely of 1.4\,s integration times is likely to be productive. While fainter sources detected exclusively in these snapshots will generally not be accessible for in-beam calibration within a typical 100\, ms baseband capture, CHIME and its Outriggers contain narrow field ($< 1$ sq deg) real-time tracking beams that can be used to continuously observe those sources over the course of their transit ($\sim $ 10 min) \citep{Aaron_tracking_beams}. Upon completion, such a campaign would significantly increase the chances of, if not guarantee, having a sufficiently nearby calibrator solution at the time of an FRB detection. A grid of compact VLBI calibrators also opens up the possibility of calibrating out additional direction-dependent systematic errors that do not have a simple analytic form and may become more relevant at higher levels of precision. For instance, while the beam phase for CHIME/FRB-Outriggers is currently not well modeled, this could in principle be calibrated out with a grid of calibrators using a facet calibration scheme similar to that developed in \cite{facet_cal}.


\subsection{Beyond CHIME/FRB Outriggers}


Our survey and future calibrator surveys with CHIME-GBO can also be used to understand the structure of AGN cores at low radio frequencies. For instance, a dedicated follow-up campaign at \SI{327}{\mega\hertz}, similar to~\citet{Solokovsky_2011}, can be used to systematically study frequency-dependent core shifts. Low-frequency VLBI observations would be most sensitive to the dense plasma environment of the AGN core.

In addition, the extended low-frequency radio emission of AGN has been shown to be correlated with the total jet power. However, measuring the extended component without the angular resolution of VLBI relies on separating the jet and core components using their different spectral indices~\citep{Fan_2018}. As a result, measurements of compact flux at low frequencies have not been available, and there is evidence that the core contribution is non-negligible at frequencies as low as 140\,MHz~\citep{Mooney_2019}. Finally, multiple groups have reported  evidence for weak positive correlation between the low-frequency radio flux and the gamma-ray flux~\citep{Giroletti_2016,Fan_2018}, suggesting a physical connection between the low-frequency and high-energy emission. In summary, dedicated follow-up of our sources will allow direct astrometric measurements of the compact core emission in blazars as well as its potential connection to high-energy emission~\citep{Plavin_2021}. 

Finally, we note that a fraction of sources in our sample are located at low Galactic latitudes. Since they are unresolved at 0.03'' scales at 600\,MHz, our measurements can potentially constrain the integrated distribution of electron fluctuations through the Galactic plane along these sightlines. In particular, the NE2001 model of Galactic angular broadening relies on a total of 94 extragalactic angular broadening measurements~\citep{Cordes_2003}. This sample approximately doubles the number of sightlines with upper limits on angular broadening, placing a useful constraints on its Milky Way contribution. This could be in turn useful for applications of interstellar optics, and in particular the scattering of FRBs~\citep{Masui_2015}. We encourage that future electron density models use our source catalog, and potential extensions thereof, to further constrain the interstellar medium within our Galaxy.


\section{Conclusion}
Using commensal observations with CHIME and its VLBI Outrigger at Green Bank Observatory, we have conducted a commensal calibrator survey covering much of the Northern sky. We have detected compact structure for $\totalsources$ sources in the RFC calibrator list, indicating that a substantial fraction of sources are unresolved at low frequencies even on 6\,M$\lambda$ scales. These sources will form an initial astrometric grid for CHIME/FRB Outriggers, an FRB survey which will precisely localize a large sample of FRBs. They are also potentially interesting targets for AGN science and modeling of the Milky Way's ISM. A few ad-hoc deeper integrations with CHIME suggest that the calibrator densities on 6\,M$\lambda$ scales at sub-GHz frequencies can be increased by a marginally deeper survey, potentially to a density sufficient for all-sky, low-frequency VLBI mapping.

\section*{Data Availability}
Our catalog is freely available for download at \url{https://zenodo.org/records/13763864}. 



\section*{Acknowledgments}
\begin{acknowledgments}
Special thanks to Dillon Dong, Craig Walker, Rick Perley, Paul Demorest, Jason Hessels, and Ziggy Pleunis for their comments and discussions on this work.  We acknowledge that CHIME is located on the traditional, ancestral, and unceded territory of the Syilx/Okanagan people.

We are grateful to the staff of the Dominion Radio
Astrophysical Observatory, which is
operated by the National
Research Council Canada. 
CHIME is funded by a grant from the Canada Foundation 
for Innovation (CFI) 2012 Leading Edge Fund (Project 31170) 
and by contributions from the provinces of British Columbia, 
Qu\'ebec and Ontario. The CHIME/FRB Project is funded by a 
grant from the CFI 2015 Innovation Fund (Project 33213) and 
by contributions from the provinces of British Columbia and 
Qu\'ebec, and by the Dunlap Institute for Astronomy and 
Astrophysics at the University of Toronto. 
Additional support was provided by the Canadian 
Institute for Advanced Research (CIFAR), McGill 
University and the McGill Space Institute via the 
Trottier Family Foundation, and the University of 
British Columbia. 
The CHIME/FRB Outriggers program is funded by 
the Gordon and Betty Moore Foundation
and by a National Science Foundation (NSF) grant (2008031).
FRB research at MIT is supported by an NSF grant (2008031).
FRB research at WVU is supported by an NSF grant (2006548, 2018490).
\end{acknowledgments}

%

\vspace{5mm}
\facilities{CHIME}


\software{\texttt{numpy} \citep{harris2020array}, \texttt{scipy} \citep{virtanen2020scipy}, \texttt{matplotlib}~\citep{hunter2007matplotlib}, 
}





\bibliography{references}{}

\bibliographystyle{aasjournal}



\end{document}